\documentclass[12pt]{article}
\usepackage{hyperref}
\usepackage[english]{babel}
\usepackage{amsmath}
\usepackage{latexsym}
\usepackage{amssymb}
\usepackage[all]{xy}
\usepackage[dvips]{graphicx}
\usepackage{epsfig}
\usepackage{psfrag}

\numberwithin{equation}{section} \numberwithin{table}{section}
\numberwithin{figure}{section}

\begin{document}



\begin{titlepage}
  \begin{flushright}
  {\small CQUeST-2010-0395}
  \end{flushright}

  \begin{center}

    \vspace{20mm}

    {\LARGE \bf Zero Sound in Effective Holographic Theories}

    \vspace{10mm}

   Bum-Hoon Lee$^{\ast\dag}$, Da-Wei Pang$^{\dag}$ and Chanyong Park$^{\dag}$

    \vspace{5mm}
    {\small \sl $\ast$ Department of Physics, Sogang University}\\
    {\small \sl Seoul 121-742, Korea\\}
    {\small \sl $\dag$ Center for Quantum Spacetime, Sogang University}\\
    {\small \sl Seoul 121-742, Korea\\}
    {\small \tt bhl@sogang.ac.kr, pangdw@sogang.ac.kr, cyong21@sogang.ac.kr}
    \vspace{10mm}

  \end{center}

\begin{abstract}
\baselineskip=18pt We investigate zero sound in $D$-dimensional effective holographic theories, whose
action is given by Einstein-Maxwell-Dilaton terms. The bulk spacetimes include both zero temperature
backgrounds with anisotropic scaling symmetry and their near-extremal counterparts obtained in 1006.2124
[hep-th], while the massless charge carriers are described by probe D-branes. We discuss thermodynamics
of the probe D-branes analytically. In particular, we clarify the conditions under which the specific
heat is linear in the temperature, which is a characteristic feature of Fermi liquids. We also compute
the retarded Green's functions in the limit of low frequency and low momentum and find quasi-particle
excitations in certain regime of the parameters. The retarded Green's functions are plotted at specific
values of parameters in $D=4$, where the specific heat is linear in the temperature and the quasi-particle
excitation exists. We also calculate the AC conductivity in $D$-dimensions as a by-product.

\end{abstract}
\setcounter{page}{0}
\end{titlepage}

\pagestyle{plain} \baselineskip=19pt

\tableofcontents

\section{Introduction}
The AdS/CFT correspondence~\cite{Maldacena:1997re, Aharony:1999ti}
has revealed the deep relations between gauge theories and
string theories and has provided powerful tools for understanding the dynamics of strongly coupled
field theories in the dual gravity side. In recent years, this paradigm has been applied to investigate
the properties of certain condensed matter systems~\cite{Hartnoll:2009sz}. The correspondence between
gravity theories and condensed matter physics(sometimes is also named as AdS/CMT correspondence)
has shed light on studying physics in the real world in the context of holography.

It is well known that in realistic condensed matter systems, the presence
of a finite density of charge carriers is of great importance. According to the AdS/CFT correspondence,
the dual bulk gravitational background should be charged black holes in asymptotically AdS spacetimes.
The simplest example of such charged AdS black holes is Reissner-Nordstr\"{o}m-AdS(RN-AdS) black hole,
which has proven to be an efficient laboratory for studying the AdS/CMT correspondence. For instance,
investigations of the fermionic two-point functions in this background indicated the existence of fermionic
quasi-particles with non-Fermi liquid behavior~\cite{Lee:2008xf, Liu:2009dm, Cubrovic:2009ye}, while the
$AdS_{2}$ symmetry of the extremal RN-AdS black hole is crucial to the emergent scaling symmetry
at zero temperature~\cite{Faulkner:2009wj}. Moreover, adding a charged scalar in such background leads to superconductivity~\cite{Gubser:2008px, Hartnoll:2008vx, Hartnoll:2008kx}.

A further step towards a holographic model-building of strongly-coupled systems at
finite charge density is to consider the leading relevant (scalar) operator in the field theory side,
whose bulk gravity theory is an Einstein-Maxwell-Dilaton system with a scalar potential.
Such theories at zero charge density were analyzed in detail in recent years as they mimic
certain essential properties of QCD~\cite{Gursoy:2007cb, Gursoy:2007er, Gubser:2008yx, Gursoy:2008bu,
Gursoy:2008za}. Solutions at finite charge density have been considered in~\cite{Gubser:2009qt, Goldstein:2009cv, Gauntlett:2009bh, Cadoni:2009xm, Chen:2010kn, Goldstein:2010aw}
in the context of AdS/CMT correspondence.

Recently a general framework for the discussion of the holographic dynamics of
Einstein-Maxwell-Dilaton systems with a scalar potential was proposed in~\cite{Charmousis:2010zz},
which was a phenomenological approach based on the concept of Effective Holographic Theory (EHT).
The minimal set of bulk fields contains the metric $g_{\mu\nu}$, the gauge field $\mathcal{A}_{\mu}$
and the scalar $\phi$ (dual to the relevant operator). $\phi$ appears in two scalar functions that
enter the effective action: the scalar potential and the non-minimal Maxwell coupling. They studied
thermodynamics of certain exact solutions and computed the DC and AC conductivity. The main advantage
of this EHT approach is that it permits a parametrization of large classes of IR dynamics and allows
investigations on important observarables. However, it is not clear whether concrete EHTs can be embedded into string theories. For subsequent generalizations see~\cite{Lee:2010xx, Lee:2010ii, Liu:2010ka,
Gursoy:2010kw, Faulkner:2010gj, Bayntun:2010nx, AliAkbari:2010av, Astefanesei:2010dk}.

On the other hand, strongly coupled quantum liquids play an important role in condensed matter
physics, where quantum liquids mean translationally invariant systems at zero (or low) temperature
and at finite density. By now there are two successful phenomenological theories of quantum liquids: Landau's
Fermi-liquid theory and the theory of quantum Bose liquids, describing two different behaviors of a quantum
liquid at low momenta and temperatures. In particular, the specific heat of a Bose liquid at low temperature
is proportional to $T^{q}$ in $q$ spatial dimensions, while the specific heat of a Fermi liquid scales
as $T$ at low $T$, irrespective of the spatial dimensions.

One may wonder if the newly developed techniques in AdS/CFT correspondence can help us understand the behavior
of quantum liquids. In~\cite{Karch:2008fa} the authors considered a class of gauge theories with fundamental
fields whose holographic dual in the appropriate limit was given in terms of the Dirac-Born-Infeld (DBI) action
in AdS space. They found that the specific heat$\sim T^{2p}$ in $p$ spatial dimensions at low temperature
and the system supported a sound mode at zero temperature, which was called ``zero-temperature sound''.
One interesting feature was that the ``holographic zero sound'' mode was almost identical to the zero sound
in Fermi liquids: the real part of the dispersion relation was linear in momentum ($\omega=qv$) and the imaginary
part had the same $q^{2}$ dependence predicted by Landau. The crucial difference was that the zero-temperature
sound velocity coincided with the first-sound velocity, while generically the two velocities are not equal
for a Fermi liquid. Such analysis was performed in the case of massive charge carriers in~\cite{Kulaxizi:2008kv} and in the case of Sakai-Sugimoto model in~\cite{Kulaxizi:2008jx}. The specific heat of general $Dp/Dq$ systems was calculated in~\cite{Karch:2009eb} and the specific heat of Lifshitz black holes
was discussed in~\cite{Lee:2010uy} and~\cite{HoyosBadajoz:2010kd}, while the zero sound was also investigated in
~\cite{HoyosBadajoz:2010kd}.

In this paper we will study the low-temperature specific heat and the holographic zero sound in effective
holographic theories. Here the bulk effective theory is $D$-dimensional Einstein gravity coupled to a Maxwell term
with non-minimal coupling and a scalar. It was found in~~\cite{Perlmutter:2010qu} that the theory admitted both extremal and near-extremal solutions with anisotropic scaling symmetry. We consider dynamics of probe
D-branes in the above mentioned backgrounds and find that by appropriately fixing the parameters in
the effective theory, the specific heat can be proportional to $T$, resembling a Fermi liquid. We also compute
the current-current retarded Green functions at low frequency and low momentum, and clarify the conditions when
a quasi-particle excitation exists. Moreover, we also explore the possibility of observing the existence of Fermi surfaces in such a system by numerical methods. We find that although the system possesses some features of Fermi liquids, such as linear specific heat and zero sound excitation, we do not observe any characteristic
structure in the wide range of $k$. In addition, the AC conductivity is also obtained as a by-product.

The rest of the paper is organized as follows: the exact solutions of the effective bulk theory
will be reviewed in section 2 and the thermodynamics of massless charge carriers will be discussed in
section 3. We shall calculate the correlation functions in section 4 and identify the quasi-particle
behavior, while the existence of Fermi surfaces will be explored in section 5 via numerics. We will calculate
the AC conductivity in section 6, including the zero density limit. Finally we will give a summary and discuss
future directions.

\section{The solution}
In this section we will review the solutions obtained in~\cite{Perlmutter:2010qu}, which can be seen
as generalizations of the four-dimensional near-extremal scaling solution discussed in~\cite{Charmousis:2010zz}.
In the beginning we consider the following action in $D$-dimensions, without any reference to string theory
or M/theory origin nor specifying the forms of the gauge coupling $f(\phi)$ and the scalar potential
$\mathcal{V}(\phi)$ explicitly,
\begin{equation}
S=-\frac{1}{16\pi G_{D}}\int d^{D}x\sqrt{-g}[R
+f(\phi)\mathcal{F}_{\mu\nu}\mathcal{F}^{\mu\nu}+\frac{1}{2}(\partial\phi)^{2}
+\mathcal{V}(\phi)].
\end{equation}
The resulting solutions are charged dilaton black holes, which have been investigated in the literature
for a long period~\cite{Gibbons:1987ps, Preskill:1991tb, Garfinkle:1990qj, Holzhey:1991bx}.
Let us focus on solutions carrying electric charge only. The general configuration with
planar symmetry can be written as follows
\begin{eqnarray}
ds^{2}&=&-U(r)dt^{2}+\frac{dr^{2}}{U(r)}+V(r)\sum\limits^{D-2}_{i=1}dx^{2}_{i},\nonumber\\
\phi&=&\phi(r),~~~\mathcal{A}_{t}=\mathcal{A}_{t}(r),~~~\mathcal{A}_{r}=\mathcal{A}_{i}=0.
\end{eqnarray}
After plugging in the scaling ansatz
\begin{equation}
U(r)\sim r^{\beta},~~~V(r)\sim r^{\gamma}
\end{equation}
into the equations of motion, we can arrive at several constraints on the parameters and the scalar functions:
\begin{itemize}
\item We require that $\beta>1$, so that the extremal solutions have smooth connections to the finite
temperature solutions;
\item The field equations indicate that $\beta\leq2$ and $0\leq\gamma\leq2$ and $\beta=2$
when the scale invariance is restored.
\item The equations of motion determine that the scalar field must take the form
\begin{equation}
\phi(r)=C_{2}\log r+\phi_{0},
\end{equation}
and both $f(\phi)$ and $\mathcal{V}(\phi)$ are constrained to be exponential in $\phi$,
power law in $r$.
\item Once we have fixed $\beta>1$, the metric
must have $\beta\geq\gamma$, where saturation occurs for vanishing flux, e.g. in $AdS_{D}$ with
$\beta=\gamma=2$.
\end{itemize}

Subsequently, according to the constraints discussed above, we take the following forms of $f(\phi)$ and $\mathcal{V}(\phi)$.
\begin{equation}
f(\phi)=e^{\alpha\phi},~~~\mathcal{V}(\phi)=-V_{0}e^{\eta\phi}.
\end{equation}
Then we will consider charged dilaton black holes with a Liouville potential~\cite{Cai:1996eg}.  Now the scaling ansatz turns out to be
\begin{eqnarray}
ds^{2}&=&-C_{1}r^{\beta}dt^{2}+\frac{dr^{2}}{C_{1}r^{\beta}}+C_{3}r^{\gamma}
\sum\limits^{D-2}_{i=1}dx^{2}_{i},\nonumber\\
\phi(r)&=&C_{2}\log r+\phi_{0},~~~\mathcal{A}^{\prime}(r)=\frac{Q}
{r^{\alpha C_{2}+\gamma\frac{D-2}{2}}}.
\end{eqnarray}
Since $\phi_{0}$ and $C_{3}$ can be eliminated by rescaling $r$ and $x_{i}$,
we shall set $\phi_{0}=0$ and $C_{3}=1$. The remaining parameters can be explicitly
given in terms of $\{V_{0}, \eta, \alpha\}$,
\begin{eqnarray}
\beta&=&2-\frac{2(D-2)(\alpha+\eta)}{(\alpha+\eta)^{2}+2(D-2)}\eta,~~~
\gamma=\frac{2(\alpha+\eta)^{2}}{(\alpha+\eta)^{2}+2(D-2)},\nonumber\\
C_{2}&=&-\frac{(D-2)}{\alpha+\eta}\gamma,~~~Q^{2}=\frac{V_{0}}{2}
\frac{2-\eta^{2}-\alpha\eta}{2+\alpha^{2}+\alpha\eta},\nonumber\\
C_{1}&=&\frac{V_{0}[(\alpha+\eta)^2+2(D-2)]^{2}}{(D-2)(2+\alpha^{2}+\alpha\eta)
[2(D-2)+(D-1)\alpha^{2}-\eta^{2}(D-3)+2\alpha\eta]}.
\end{eqnarray}
It can be easily seen that there is no scale invariance in such backgrounds.
We will restrict to $\eta>0$ and $V_{0}>0$ without loss of generality, which implies that
$\phi$ must diverge to positive infinity at the horizon.

In the specific limit $\eta=0, \beta=2$, the scalar potential becomes constant and the
scale invariance of the solutions can be restored:
\begin{itemize}
\item When we set $\alpha\rightarrow+\infty, \gamma=2$ with vanishing flux, the scalar $\phi$
becomes constant and the resulting solution is $AdS_{D}$;
\item When we set $\alpha~\rightarrow~0, \gamma=0$ with flux through
$\textbf{R}^{D-2}$, $\phi$ becomes constant and the resulting solution is $AdS_{2}\times\textbf{R}^{D-2}$;
\item When we set $\alpha$ arbitrary, $0\leq\gamma\leq2$ with flux through
$\textbf{R}^{D-2}$, the scalar $\phi\sim\log r$ and the resulting solution is the modified Lifshitz solution,  whose dynamical exponent $\textsf{z}=2/\gamma$\footnote{The reason why such a solution is called ``modified Lifshitz'' is that the scalar field must be constant in Lifshitz background, which is required by scaling symmetry~\cite{Kachru:2008yh}. Properties of the modified Lifshitz solutions have been studied in~\cite{Taylor:2008tg} and~\cite{Chen:2010kn}.}
\end{itemize}

Before coming to practical calculations we should determine the range of parameters. Firstly, by requiring that $\phi(r)~\rightarrow~+\infty$ for small $r$ and the flux to be real,
one can impose the bound on $\alpha$ in terms of fixed $\eta$,
\begin{equation}
-\eta<\alpha<\frac{2}{\eta}-\eta.
\end{equation}
Secondly, when the flux is zero,
$$\alpha=\frac{2}{\eta}-\eta,~~~\beta=\gamma=\frac{4}{2+(D-2)\eta^{2}}\equiv\gamma^{\prime}.$$
we should require $\gamma^{\prime}>1$ to ensure a well-defined boundary in the sense of AdS/CFT
\footnote{For details see~\cite{Perlmutter:2010qu}.},
$$
\eta^{2}<\frac{2}{D-2}.
$$
Combining constraint derived in the general background in the beginning of this section, we can obtain the following complete restrictions
\begin{equation}
\label{res}
1<\beta\leq2,~~\gamma\leq\beta\leq2,~~-\eta<\alpha\leq\frac{2}{\eta}-\eta,~~0\leq\eta<\sqrt{\frac{2}{D-2}}.
\end{equation}
We will impose such constraints in the subsequent calculations.

The near extremal solution can be obtained in a similar way,
\begin{equation}
ds^{2}=-C_{1}r^{\beta}f(r)dt^{2}+\frac{dr^{2}}{C_{1}r^{\beta}f(r)}+C_{3}r^{\gamma}
\sum\limits^{D-2}_{i=1}dx^{2}_{i},
\end{equation}
where
\begin{equation}
f(r)=1-(\frac{r_{+}}{r})^{w},~~~w=\beta-1+\frac{D-2}{2}\gamma,
\end{equation}
and the other parameters and fields remain the same as the extremal solutions.
One can easily get the temperature
\begin{equation}
T=\frac{1}{4\pi}C_{1}wr^{\beta-1}_{+},
\end{equation}
and the entropy density
\begin{equation}
s\equiv\frac{S_{BH}}{V_{\textbf{R}^{D-2}}}=\frac{1}{4G_{D}}r_{+}^{\frac{D-2}{2}\gamma}.
\end{equation}

\section{Thermodynamics of probe D-branes}
In this section we will investigate thermodynamics of massless charge carriers in the backgrounds
reviewed in previous section. According to AdS/CFT, $N_{f}$ probe D-branes correspond to $N_{f}$
fields in the fundamental representation of the gauge group in the probe limit $N_{f}\ll N$~\cite{Karch:2002sh, Kobayashi:2006sb}. An efficient method for evaluating the DC conductivity and DC Hall conductivity of
probe D-branes was proposed in~\cite{Karch:2007pd} and~\cite{O'Bannon:2007in}. Moreover, a holographic model
building approach to ``strange metallic'' phenomenology was initiated in~\cite{Hartnoll:2009ns}, where the bulk
spacetime was a Lifshitz black hole and the charge carriers were described by D-branes. Here we will consider
probe D-branes as massless charge carriers and explore the thermodynamics in the near-extremal background.

The dynamics of probe D-branes is described by the Dirac-Born-Infeld (DBI) action
\begin{eqnarray}
S_{\rm DBI}&=&-N_{f}T_{D}{\rm Vol(\Sigma)}\int dtdrd^{q}xe^{-\phi}
\sqrt{-{\rm det}(g_{ab}+2\pi\alpha^{\prime}F_{ab})},\nonumber\\
&=&-\tau_{\rm eff}\int dtdrd^{q}xe^{-\phi}
\sqrt{-{\rm det}(g_{ab}+2\pi\alpha^{\prime}F_{ab})},
\end{eqnarray}
where $T_{D}$ denotes the tension of D-branes, $g_{ab}$ is the induced metric and $F_{ab}$ is the
$U(1)$ field strength on the worldvolume. In the second line we set $\tau_{\rm eff}=N_{f}T_{D}{\rm Vol(\Sigma)}$, where ${\rm Vol}(\Sigma)$ denotes volume of the internal space that the D-branes may be wrapping. Furthermore,
we assume that the D-branes are extended along $q\leq D-2$ spatial dimensions of the black hole solution.
If $q<D-2$, the fundamental fields are propagating along certain $q$-dimensional defect. We will
introduce a nontrivial worldvolume gauge field $A_{t}(r)$ and absorb the
factor $2\pi\alpha^{\prime}$ into $F_{ab}$. Since we are not studying realistic string theories, the Wess-Zumino terms will be omitted in the following discussions.

Before proceeding we should make sure that the backreaction of the probe branes onto
the background can be neglected. Our discussion is along the line of~\cite{Hartnoll:2009ns}.
Expanding the DBI action to quadric order of $F_{rt}$ in the background, we can obtain
\begin{equation}
S_{\rm DBI}=-\tau_{\rm eff}\int dtdrd^{q}xe^{-\phi}\sqrt{-g}\sqrt{1+g^{tt}g^{rr}F^{2}_{rt}},
\end{equation}
To avoid backreaction of the probes on the background, the stress energy of the probes
must be smaller than that generating the bulk spacetime. It can be easily seen that the stress
energy of the original background$\sim\ell_{P}^{D-2}|\Lambda|$, where $\ell_{P}$ denotes the Planck length
in $D$-dimensional spacetime and $\Lambda$ is the corresponding cosmological constant. Therefore by varying the
quadric action of the probes with respect to $g_{tt}$, we can arrive at the following condition
\begin{equation}
\frac{e^{-\phi}}{\sqrt{1+g^{tt}g^{rr}F^{2}_{rt}}}
\ll\frac{\ell_{P}^{D-2}|\Lambda|}{\tau_{\rm eff}}.
\end{equation}
One can see that as long as the effective tension $\tau_{\rm eff}$ is sufficiently small,
the backreaction can be neglected.

In this background configuration, after performing the trivial
integrations on $dtd^{q}x$ and dividing out the infinite volume of
$\textbf{R}^{1,q}$, we can obtain the action density.
\begin{equation}
S=-\tau_{\rm eff}\int drr^{m}\sqrt{1-A^{\prime2}_{t}},
\end{equation}
where
\begin{equation}
m=\frac{1}{2}\gamma q-C_{2}=\frac{(\alpha+\eta)
[q(\alpha+\eta)+2(D-2)}{(\alpha+\eta)^{2}+2(D-2)}.
\end{equation}
and the prime denotes derivative with respect to $r$. Then the charge density
is given by
\begin{equation}
\rho\equiv\frac{\delta\mathcal{L}}{\delta A^{\prime}_{t}}=\tau_{\rm eff}
\frac{r^{m}A^{\prime}_{t}}{\sqrt{1-A^{\prime2}_{t}}}.
\end{equation}
We can also solve for $A^{\prime}_{t}(r)$
\begin{equation}
A^{\prime}_{t}=\frac{d}{\sqrt{r^{2m}+d^{2}}},~~~d\equiv\frac{\rho}{\tau_{\rm eff}}.
\end{equation}
By plugging in the solution for $A^{\prime}_{t}(r)$, we can find the on-shell action
density
\begin{equation}
S_{\rm on-shell}=-\tau_{\rm eff}\int dr\frac{r^{2m}}{\sqrt{r^{2m}+d^{2}}}.
\end{equation}

Following the methods used in~\cite{Karch:2007br}, some interesting physical quantities
like the chemical potential and the free energy, can be evaluated analytically.
We will see that this is still the case for our background. The chemical potential is given by
\begin{eqnarray}
\mu&=&\int^{\infty}_{r_{+}}A^{\prime}_{t}dr,\nonumber\\
&=&\mu_{0}-r_{+}{}_{2}F_{1}(\frac{1}{2m},\frac{1}{2};1+\frac{1}{2m},-\frac{r^{2m}_{+}}{d^{2}}),
\end{eqnarray}
where
\begin{equation}
\mu_{0}=d^{\frac{1}{m}}B_{0}(m),~~~B_{0}(m)=\frac{1}{2}B(1+\frac{1}{2m},
\frac{1}{2}-\frac{1}{2m}).
\end{equation}
Notice that in order to obtain the above results, we
have made use of the following useful formulae for Beta function and incomplete Beta function
\begin{eqnarray*}
B(a,b)&=&\frac{\Gamma(a)\Gamma(b)}{\Gamma(a+b)}=\int^{\infty}_{0}du(1+u)^{-(a+b)}u^{a-1},\nonumber\\
B(x;a,b)&=&\int^{x/(1-x)}_{0}du(1+u)^{-(a+b)}u^{a-1}.
\end{eqnarray*}
as well as for Hypergeometric function
\begin{eqnarray*}
B(x;a,b)&=&a^{-1}{x^{a}}_{2}F_{1}(a,1-b;a+1,x),\nonumber\\{}_{2}F_{1}(a,b;c,x)&=&
(1-x)^{-a}{}_{2}F_{1}(a,c-b;c,\frac{x}{x-1}).
\end{eqnarray*}
After choosing the grand-canonical ensemble, the free energy density is given by
\begin{eqnarray}
\Omega&=&-S_{\rm on-shell}=\tau_{\rm eff}\int^{\infty}_{r_{+}}dr\frac{r^{2m}}{\sqrt{r^{2m}+d^{2}}},\nonumber\\
&=&\Omega_{0}-\frac{\tau_{\rm eff}}{(2m+1)d}r^{2m+1}_{+}{}_
{2}F_{1}(1+\frac{1}{2m},\frac{1}{2};2+\frac{1}{2m},-\frac{r^{2m}_{+}}{d^{2}}),
\end{eqnarray}
where
\begin{equation}
\Omega_{0}=-\frac{\tau_{\rm eff}}{2(m+1)}d^{1+\frac{1}{m}}B(1+\frac{1}{2m},\frac{1}{2}-\frac{1}{2m})
=-\frac{\tau_{\rm eff}}{(m+1)B_{0}(m)^{m}}\mu_{0}^{m+1}.
\end{equation}

Moreover, other thermodynamic quantities can also be calculated from the thermodynamic relations.
The charge density can be written as
\begin{equation}
\rho=-\frac{\partial\Omega}{\partial\mu}=\tau_{\rm eff}(\frac{\mu_{0}}{B_{0}(m)})^{m}=\tau_{\rm eff}d,
\end{equation}
which is consistent with previous result.
The entropy density is given by
\begin{equation}
s=-\frac{\partial\Omega}{\partial T}=\frac{\rho}{\beta-1}(\frac{4\pi}{C_{1}w})^{\frac{1}{\beta-1}}
T^{\frac{2-\beta}{\beta-1}}+\frac{\tau_{\rm eff}}{2(\beta-1)d}(\frac{4\pi}{C_{1}w})^{\frac{2m+1}{\beta-1}}
T^{\frac{2(m+1)-\beta}{\beta-1}},
\end{equation}
Notice that when $\beta=2$, there exists a nontrivial contribution
to the entropy density at $T=0$ like those observed in~\cite{Lee:2010uy} and~\cite{HoyosBadajoz:2010kd}.
On the other hand, the entropy density is vanishing at extremality as long as $\beta\neq2$.
The specific heat is
\begin{equation}
c_{V}=T\frac{\partial s}{\partial T}=\frac{\rho(2-\beta)}{(\beta-1)^{2}}(\frac{4\pi}{C_{1}w})^{\frac{1}{\beta-1}}
T^{\frac{2-\beta}{\beta-1}}+\frac{\tau_{\rm eff}^{2}(2m+2-\beta)}{2(\beta-1)^{2}\rho}
(\frac{4\pi}{C_{1}w})^{\frac{2m+1}{\beta-1}}T^{\frac{2m+2-\beta}{\beta-1}}.
\end{equation}
It is well known that for a gas of free bosons in $q$ spatial dimensions, the specific
heat at low temperature is proportional to $T^{p}$, while for a gas of fermions the low
temperature specific heat is proportional to $T$, irrespective of $p$. When
$\beta\neq2$ and $\tau_{\rm eff}$ is sufficiently small, the first term dominates.
One can easily obtain $\beta=3/2$ when the specific heat is proportional to $T$. Then the
parameter $\alpha$ can be expressed in terms of $\eta$
\begin{equation}
\alpha_{1\pm}=(2D-5)\eta\pm\sqrt{2(D-2)[2(D-2)\eta^{2}-1]},
\end{equation}
Combining with~(\ref{res}), we can arrive at the following conclusions:
\begin{itemize}
\item When
$$\sqrt{\frac{1}{2(D-2)}}\leq\eta<\sqrt{\frac{2}{3(D-2)}},$$
both $\alpha_{1+}$ and $\alpha_{1-}$ are permitted solutions;
\item When $$\sqrt{\frac{2}{3(D-2)}}\leq\eta<\sqrt{\frac{2}{D-2}},$$
only $\alpha_{1-}$ is a permitted solution;
\item When $$0<\eta<\sqrt{\frac{1}{2(D-2)}},$$ there is no solution, which means
that we cannot realize $c_{V}\propto T$ in this regime.
\end{itemize}
When $\beta=2, \eta=0$, the second term provides the only contribution. The
linear dependence on $T$ fixes $m=1/2$. Note that in this limit, $\gamma=2\alpha^{2}/(\alpha^{2}+2(D-2))\equiv2/\textsf{z}$.
So
\begin{equation}
\alpha_{2\pm}=\frac{-4(D-2)\pm\sqrt{16(D-2)^{2}+4(D-2)(2q-1)}}{2(2q-1)},
\end{equation}
By taking into account of~(\ref{res}), we can find that only $\alpha_{2+}$ is
permitted.

We can also formally evaluate the ``speed of sound''. In grand-canonical ensemble, the
pressure is given by
\begin{equation}
P=-\Omega_{0}=\frac{\tau_{\rm eff}}{(m+1)B_{0}(m)^{m}}\mu_{0}^{m+1},
\end{equation}
while the energy density is
\begin{equation}
\varepsilon=\Omega_{0}+\mu_{0}\rho=\frac{m\tau_{\rm eff}}{(m+1)B_{0}(m)^{m}}\mu_{0}^{m+1}.
\end{equation}
Therefore,
\begin{equation}
\varepsilon=mP,~~~\Rightarrow~~~c_{s}^{2}=\frac{\partial P}{\partial\varepsilon}=\frac{1}{m}.
\end{equation}
However, it was emphasized in~\cite{HoyosBadajoz:2010kd} that
this quantity is only the speed of normal/first sound in the relativistic case $\textsf{z}=1$.
Actually the speed of normal sound is dimensionful in a system with $\textsf{z}>1$.
We will calculate the holographic zero sound in the next section.

\section{The holographic zero sound}
In this section we will calculate the holographic zero sound in the anisotropic background
at extremality. The basic strategy is to consider fluctuations of the worldvolume gauge field
on the probe D-branes in the background with nontrivial $A_{t}$. Such analysis was performed for
$AdS_{p+2}$ background in~\cite{Karch:2008fa} and for Lifshitz background in~\cite{HoyosBadajoz:2010kd}.
We will calculate the holographic zero sound in a similar way and classify the behavior of the zero sound
in different parameter ranges.
\subsection{The retarded Green's functions}
Zero sound should appear as a pole in the density-density retarded two-point
function $G^{R}_{tt}(\omega, k)$ at extremality~\cite{Karch:2008fa}.
In~\cite{HoyosBadajoz:2010kd} the authors provided a general framework for evaluating the
corresponding retarded Green's functions with background metric
\begin{equation}
ds^{2}=g_{tt}dt^{2}+g_{rr}dr^{2}+g_{xx}\sum\limits^{D-2}_{i=1}dx^{2}_{i}.
\end{equation}
Here we will take the nontrivial dilaton into account. The symmetries in the spatial directions
allow us to consider fluctuations of the gauge fields with the following form $$A_{\mu}(r)~\rightarrow~A_{\mu}(r)+a_{\mu}(t,r,x),$$
where $x$ denotes one of the spatial directions.
The quadratic action for the fluctuations is given by
\begin{equation}
S_{a^{2}}=\frac{\tau_{\rm eff}}{2}\int dtdrd^{q}xe^{-\phi}g^{q/2}_{xx}
[\frac{g_{rr}f^{2}_{tx}-|g_{tt}|a^{\prime2}_{x}}{g_{xx}(|g_{tt}|g_{rr}-A^{\prime2}_{t})^{1/2}}+
\frac{|g_{tt}|g_{rr}a^{\prime2}_{t}}{(|g_{tt}|g_{rr}-A^{\prime2}_{t})^{3/2}}],
\end{equation}
where $f_{tx}=\partial_{t}a_{x}-\partial_{x}a_{t}$. Note that we are working
in the gauge of $a_{r}=0$. After performing the Fourier transform
$$a_{\mu}(t,r,x)=\int\frac{d\omega dk}{(2\pi)^{2}}e^{-i\omega t+ikx}a_{\mu}(\omega,r,k),$$
the linearized equations of motion can be written as
\begin{equation}
\partial_{r}[\frac{e^{-\phi} g_{xx}^{q/2}|g_{tt}|g_{rr}a_{t}^{\prime}}
{(|g_{tt}|g_{rr}-A^{\prime2}_{t})^{3/2}}]-\frac{e^{-\phi}g^{q/2-1}_{xx}g_{rr}}
{\sqrt{|g_{tt}|g_{rr}-A^{\prime2}_{t}}}(k^{2}a_{t}+\omega ka_{x})=0,
\end{equation}
\begin{equation}
\partial_{r}[\frac{e^{-\phi}g_{xx}^{q/2-1}|g_{tt}|a_{x}^{\prime}}
{(|g_{tt}|g_{rr}-A^{\prime2}_{t})^{1/2}}]+\frac{e^{-\phi}g^{q/2-1}_{xx}g_{rr}}
{\sqrt{|g_{tt}|g_{rr}-A^{\prime2}_{t}}}(\omega^{2}a_{x}+\omega ka_{t})=0.
\end{equation}
In addition, the following constraint can be obtained by writing $a_{r}$'s equation of
motion in $a_{r}=0$ gauge
\begin{equation}
g_{rr}g_{xx}\omega a^{\prime}_{t}+(|g_{tt}|g_{rr}-A^{\prime2}_{t})ka^{\prime}_{x}=0.
\end{equation}
The above equations are not independent, as we can obtain the equation
of motion for $a_{t}$ by combining the constraint equation and the equation of motion for $a_{x}$.
Therefore it is sufficient to solve the constraint and the equation for $a_{t}$ only.
By introducing the gauge-invariant electric field
$$E(r,\omega,k)=\omega a_{x}+ka_{t},$$  we can obtain the equation of motion for $E$
\begin{equation}
\label{4eq6}
E^{\prime\prime}+[\partial_{r}\ln(\frac{e^{-\phi}g^{\frac{q-3}{2}}_{xx}|g_{tt}|g_{rr}^{-1/2}}
{u(k^{2}u^{2}-\omega^{2})})]E^{\prime}-\frac{g_{rr}}{|g_{tt}|}(u^{2}k^{2}-\omega^{2})E=0,
\end{equation}
where
\begin{equation}
u(r)=\sqrt{\frac{|g_{tt}|g_{rr}-A^{\prime2}_{t}}{g_{rr}g_{xx}}}.
\end{equation}
Moreover, the quadratic action can also be expressed in terms of $E$,
\begin{equation}
S_{a^{2}}=\frac{\tau_{\rm eff}}{2}\int drd\omega dk\frac{e^{-\phi}g^{\frac{q-3}{2}}g_{rr}^{1/2}}
{u}[E^{2}+\frac{|g_{tt}|}{g_{rr}(u^{2}k^{2}-\omega^{2})}E^{\prime2}],
\end{equation}

For our specific background, the metric can be rewritten in terms of the new radial coordinate
$z=1/r$ as follows
\begin{equation}
ds^{2}=-\frac{C_{1}}{z^{\beta}}dt^{2}+\frac{z^{\beta-4}}{C_{1}}dz^{2}
+\frac{1}{z^{\gamma}}\sum\limits^{D-2}_{i=1}dx^{2}_{i}.
\end{equation}
In the new coordinate system, the solution of the worldvolume gauge field and
the function $u(z)$ are given by
\begin{equation}
\dot{A}_{t}^{2}=\frac{d^{2}z^{2m}}{z^{4}(1+d^{2}z^{2m})},~~~
u^{2}=\frac{C_{1}}{z^{\beta-\gamma}(1+d^{2}z^{2m})},
\end{equation}
where dot denotes derivative with respect to $z$.
Integrating the quadratic action by parts
\begin{equation}
S_{a^{2}}=\frac{\tau_{\rm eff}}{2}\int d\omega dk\frac{e^{-\phi}g^{\frac{q-3}{2}}_{xx}
g_{rr}^{-1/2}|g_{tt}|}{u(u^{2}k^{2}-\omega^{2})}\dot{E}E,
\end{equation}
introducing a cutoff at $z=\epsilon$ and taking the limit $\epsilon\rightarrow0$, the quadratic
action turns out to be
\begin{equation}
S_{a^{2}}=-\frac{\tau_{\rm eff}}{2}\int d\omega dk\frac{\epsilon^{2-m}}{k^{2}}\dot{E}E.
\end{equation}

After imposing the incoming boundary condition at the ``horizon'' $z\rightarrow0$ and
plugging in the solutions of $a_{\mu}$, the retarded correlation function reads~\cite{Son:2002sd}
\begin{equation}
G^{R}_{tt}(\omega,k)=\frac{\delta^{2}}{\delta a_{t}(\epsilon)^{2}}S_{a^{2}}=
\big(\frac{\delta E(\epsilon)}{\delta a_{t}(\epsilon)}\big)^{2}\frac{\delta^{2}}{\delta E(\epsilon)^{2}}S_{a^{2}},
\end{equation}
By defining
\begin{equation}
\Pi(\omega, k)\equiv\frac{\delta^{2}}{\delta E(\epsilon)^{2}}S_{a^{2}},
\end{equation}
the retarded correlation functions can be written in terms of $\Pi(\omega,k)$
\begin{equation}
G^{R}_{tt}(\omega,k)=k^{2}\Pi(\omega,k),~~~G^{R}_{tx}(\omega,k)=\omega k\Pi(\omega,k),
~~~G^{R}_{xx}(\omega,k)=\omega^{2}\Pi(\omega,k).
\end{equation}
\subsection{Matching the solutions}
In order to evaluate the retarded correlation functions, we should try to solve~(\ref{4eq6}),
whose analytic solutions are always difficult to find. We will leave the numerical
work to section 5, while here we will obtain the low-frequency
behavior of $\Pi(\omega, k)$ by solving~(\ref{4eq6}) in different limits and matching the two solutions
in an overlapped regime, following the spirit of~\cite{Karch:2008fa} and~\cite{HoyosBadajoz:2010kd}.
To be concrete, we will solve~(\ref{4eq6}) in the limit of large $z$ and then expand the solution
in the small frequency and momentum limit. Next we will take the small frequency and momentum limit
first and then perform the large $z$ expansion. The integration constants can be fixed by matching
the two solutions.

First let us take $z\rightarrow\infty$, which leads to the following equation for $E$
\begin{equation}
\ddot{E}+\frac{2-\beta+\gamma}{z}\dot{E}+\frac{\omega^{2}z^{2\beta-4}}{C^{2}_{1}}E=0.
\end{equation}
The solution can be given in terms of a Hankel function of the first kind,
\begin{equation}
\label{daz}
E=D_{0}(\frac{x}{2})^{\nu}H^{(1)}_{\nu}(x),~~~x=\frac{\omega}{C_{1}(\beta-1)}z^{\beta-1},
~~~\nu=\frac{1}{2}-\frac{\gamma}{2(\beta-1)},
\end{equation}
In the limit of small frequency with $\nu\neq0$, the asymptotic expansion reads
\begin{eqnarray}
E&=&D_{0}\Gamma(\frac{1}{2}+\frac{\gamma}{2(\beta-1)})^{-1}(1-i\tan\frac{\pi\gamma}{2(\beta-1)})\nonumber\\
& &-\frac{iD_{0}}{\pi}\Gamma(\frac{\gamma}{2(\beta-1)}-\frac{1}{2})
(\frac{\omega}{2C_{1}(\beta-1)})^{1-\frac{\gamma}{\beta-1}}z^{\beta-\gamma-1},
\end{eqnarray}
It should be pointed out that the case of $\nu=0$ must be treated separately.
In this case the corresponding parameter is given by
\begin{equation}
\alpha_{3+}=-(D-1)\eta+\sqrt{(D-2)^2\eta^{2}+2(D-2)}.
\end{equation}
Now the expansion contains a logarithmic term
\begin{equation}
E\simeq D_{0}+\frac{2i}{\pi}D_{0}(\log(\omega z^{\beta-1})-\log(2C_{1}(\beta-1))+\gamma_{E}),
\end{equation}
where $\gamma_{E}$ is the Euler constant.

Next we take $\omega z^{\beta/2}\ll1, kz^{\gamma/2}\ll1$ with $\omega k^{-\beta/\gamma}$ being fixed.
Then the last term in~(\ref{4eq6}) can be neglected and the equation of $E$ becomes
\begin{equation}
\label{small}
\ddot{E}+[\partial_{z}\ln(\frac{e^{-\phi}g^{\frac{q-3}{2}}_{xx}|g_{tt}|g_{rr}^{-1/2}}
{u(k^{2}u^{2}-\omega^{2})})]\dot{E}=0.
\end{equation}
When $m\neq1$ and $m\neq\gamma-\beta+1$, the solution is given by
\begin{eqnarray}
E&=&D_{1}+D_{2}\big[C_{1}k^{2}\frac{z^{m-1}}{m}\big(\frac{1}{\sqrt{1+d^{2}z^{2m}}}+\frac{1}{m-1}
{}_{2}F_{1}(\frac{1}{2},\frac{1}{2}-\frac{1}{2m};\frac{3}{2}-\frac{1}{2m},-d^{2}z^{2m})\big)\nonumber\\
& &-\omega^{2}\frac{z^{m+n}}{m+n}{}_{2}F_{1}(\frac{1}{2},\frac{1}{2}+\frac{n}{2m};\frac{3}{2}+\frac{n}{2m},
-d^{2}z^{2m})\big],
\end{eqnarray}
where $n\equiv\beta-\gamma-1$. For either $m\neq1$ or $m\neq\gamma-\beta+1$, the powers
of $z$ do not match, which will be displayed in Appendix A.
We will make use of the following useful formulae for the asymptotic expansion
\begin{eqnarray*}
{}_{2}F_{1}(a,b;c,x)&=&(1-x)^{-a}\frac{\Gamma(c)\Gamma(b-a)}{\Gamma(b)\Gamma(c-a)}
{}_{2}F_{1}(a,c-b;a-b+1,\frac{1}{1-x})\nonumber\\
& &+(1-x)^{-b}\frac{\Gamma(c)\Gamma(a-b)}{\Gamma(a)\Gamma(c-b)}
{}_{2}F_{1}(b,c-a;b-a+1,\frac{1}{1-x}),
\end{eqnarray*}
\begin{equation*}
{}_{2}F_{1}(\frac{1}{2},\frac{1}{2};\frac{3}{2},-x^{2})=x^{-1}\log(x+\sqrt{1+x^{2}}).
\end{equation*}
Therefore the large $z$ limit is given as follows when $n\neq0$,
\begin{equation}
E\simeq D_{1}+D_{2}[\frac{C_{1}\mu_{0}k^{2}}{md}-\frac{\omega^{2}z^{n}}{nd}-\frac{\omega^{2}
d^{-\frac{n}{m}}}{2md}B(\frac{1}{2}+\frac{n}{2m},-\frac{n}{2m})].
\end{equation}
When $n=0$ the expansion also contains a logarithmic term
\begin{eqnarray}
E&\simeq&D_{1}+D_{2}[\frac{C_{1}\mu_{0}k^{2}}{md}-\frac{\omega^{2}}{md}\log(2dz^{m})]\nonumber\\
&\simeq&D_{1}+D_{2}[\frac{C_{1}\mu_{0}k^{2}}{md}-\frac{\omega^{2}}{md}\log 2d
+\frac{\omega^{2}}{(\beta-1)d}\log\omega-\frac{\omega^{2}}{(\beta-1)d}\log(\omega z^{\beta-1})].
\end{eqnarray}

To evaluate the retarded correlation functions, we need small $z$ expansion of the solution.
It can be seen that the second term in the expansion always tends to zero more rapidly, being irrespective
of $n=0$ or not. Therefore we have
\begin{equation}
E\simeq D_{1}+D_{2}\frac{C_{1}k^{2}}{m-1}z^{m-1},
\end{equation}
Assuming that $m>1$\footnote{the $m<1$ can be dealt with in a similar fashion, see footnote 7 of~\cite{HoyosBadajoz:2010kd}.},
the leading order behavior of $E$ reads
$E(\epsilon)\simeq D_{1}, \dot{E}(\epsilon)\simeq D_{2}C_{1}k^{2}\epsilon^{m-2}$,
so the quadratic action turns out to be
\begin{eqnarray}
S_{a^{2}}&=&-\frac{\tau_{\rm eff}}{2}\int d\omega dk\frac{\epsilon^{2-m}}{k^{2}}E\dot{E}\nonumber\\
&=&-\frac{\tau_{\rm eff}C_{1}}{2}\int d\omega dkD_{1}D_{2},
\end{eqnarray}
Thus
\begin{equation}
\Pi(\omega,k)=\lim_{\epsilon\rightarrow0}\frac{\delta^{2}}{\delta E(\epsilon)^{2}}S_{a^{2}}
=\frac{\delta^{2}}{\delta D_{1}^{2}}S_{a^{2}}\big|_{\epsilon\rightarrow0},
\end{equation}
The relation between the integration constants $D_{1}$ and $D_{2}$ can be obtained by
matching the expansions of the solutions in different
limits and eliminating the other integration constant $D_{0}$.

Finally we summarize our result for $\Pi(\omega,k)$
\begin{equation}
\Pi(\omega,k)\propto\frac{\tau_{\rm eff}C_{1}}{\delta_{1}k^{2}-\delta_{2}\omega^{2}
-\delta_{3}G_{0}(\omega)}.
\end{equation}
When $n\neq0$ the parameters are given by
\begin{eqnarray}
\delta_{1}&=&-\frac{C_{1}\mu_{0}n}{\pi m(2C_{1}(\beta-1))^{\frac{n}{\beta-1}}}
\Gamma(\frac{\gamma}{2(\beta-1)}-\frac{1}{2})\Gamma(\frac{\gamma}{2(\beta-1)}+\frac{1}{2}),\nonumber\\
\delta_{2}&=&-\frac{nd^{-\frac{n}{m}}}{2\pi m(2C_{1}(\beta-1))^{\frac{n}{\beta-1}}}
\Gamma(\frac{\gamma}{2(\beta-1)}-\frac{1}{2})\Gamma(\frac{\gamma}{2(\beta-1)}+\frac{1}{2})
B(\frac{1}{2}+\frac{n}{2m},-\frac{n}{2m}),\nonumber\\
\delta_{3}&=&i+\tan\frac{\pi\gamma}{2(\beta-1)},~~~G_{0}(\omega)=\omega^{2-\frac{n}{\beta-1}}.
\end{eqnarray}
When $n=0$ the parameters are given by
\begin{eqnarray}
\delta_{1}&=&\frac{2C_{1}(\beta-1)\mu_{0}}{\pi m},~~~\delta_{2}=i,~~~\delta_{3}=-\frac{1}{\pi},\nonumber\\
G_{0}(\omega)&=&\omega^{2}\log(\delta\omega^{2}),~~~\delta=\frac{de^{2\gamma_{E}}}{2C^{2}_{1}(\beta-1)^{2}}.
\end{eqnarray}
\subsection{Zero sound}
The dispersion relation of the holographic sound mode is given by setting the
denominator of $\Pi(\omega,k)$ to vanish. Similar to the situations discussed in~\cite{HoyosBadajoz:2010kd},
the value of $n$ determines which term in the denominator dominates.

It can be seen that when $n<0$, the $\omega^{2}$ term dominates. Then we can expand $k(\omega)$
\begin{equation}
k=\pm\omega\sqrt{\frac{\delta_{2}}{\delta_{1}}}[1+\frac{\delta_{3}}{2\delta_{2}}
\omega^{-1+\frac{\gamma}{\beta-1}}+\mathcal{O}(\omega^{-2+\frac{2\gamma}{\beta-1}})],
\end{equation}
and then invert to find
\begin{equation}
\omega(k)=\pm k\sqrt{\frac{\delta_{1}}{\delta_{2}}}-\frac{\delta_{3}}{2\delta_{2}}
(\frac{\delta_{1}}{\delta_{2}})^{\frac{\gamma}{2(\beta-1)}}k^{\frac{\gamma}{\beta-1}}+
\mathcal{O}(k^{-1+\frac{2\gamma}{\beta-1}}).
\end{equation}
Notice that when $n<0$, $\beta-1<\gamma$, so at low momentum $k^{\frac{\gamma}{\beta-1}}<k$,
that is, the real part is bigger than the imaginary part. Therefore this mode describes
a quasi-particle excitation. As a check of consistency, we can take the specific limit
$\beta=\gamma=2, C_{1}=1, \lambda=0, m=q$, which just gives the $AdS_{D}$ background.
One can obtain
\begin{equation}
\omega(k)=\pm\frac{k}{\sqrt{q}}-i\frac{d^{-\frac{1}{q}}\Gamma(\frac{1}{2})}
{\Gamma(\frac{1}{2}-\frac{1}{2q})\Gamma(\frac{1}{2q})}k^{2}+\mathcal{O}(k^{3}),
\end{equation}
which agrees with~\cite{Karch:2008fa}. The speed of the holographic zero sound is given by
\begin{equation}
v^{2}_{0}=\frac{\delta_{1}}{\delta_{2}}=\frac{C_{1}}{m}d^{\frac{n+1}{m}}
\frac{\Gamma(\frac{1}{2m})\Gamma(\frac{1}{2}-\frac{1}{2m})}{\Gamma(\frac{1}{2}+\frac{n}{2m})\Gamma(-\frac{n}{2m})}.
\end{equation}
It can be seen that in the relativistic case $\beta=\gamma=2$,
the speed of zero sound coincides with the speed of normal/first sound.
One special case is $m+n=1$, where all the $\Gamma$ functions cancel and the speed of zero sound reads
\begin{equation}
v^{2}_{0}=\frac{C_{1}}{m}d^{\frac{n+1}{m}},
\end{equation}
which turns out to be finite as $n\rightarrow0$. When $m+n>1$ as well as $n<0, \Gamma(-\frac{n}{2m})$ has a pole as $n\rightarrow0$, so $v^{2}_{0}$ goes to zero from above.

When $n>0, G_{0}(\omega)$ dominates, then
\begin{equation}
k(\omega)=\pm\big(\frac{\delta_{3}}{\delta_{1}}\big)^{\frac{1}{2}}\omega^{1-\frac{n}{2(\beta-1)}}
(1+\frac{\delta_{2}}{2\delta_{3}}\omega^{\frac{n}{\beta-1}}+\mathcal{O}(\omega^{\frac{2n}{\beta-1}})).
\end{equation}
Inverting the relation above,
\begin{equation}
\omega=\big(\frac{\delta_{1}}{i\delta_{3}}\big)^{\frac{\beta-1}{2(\beta-1)-n}}k^{\frac{2(\beta-1)}{2(\beta-1)-n}}
+\frac{\beta-1}{2(\beta-1)-n}\frac{\delta_{2}}{\delta_{3}}
\big(\frac{\delta_{1}}{\delta_{3}}\big)^{\frac{\beta-1+n}{2(\beta-1)-n}}k^{\frac{2(\beta-1)-n}{2(\beta-1)-n}}
+\mathcal{O}(k^{\frac{2(\beta-1)+4n}{2(\beta-1)-n}}),
\end{equation}
Notice that since $\delta_{1}$ is real and $\delta_{3}$ is complex, the leading term has a complex coefficient.
Furthermore, the real and imaginary parts are of the same order, hence this mode is
not a quasi-particle.

Finally when $n=0$,
\begin{equation}
k=\pm\frac{\omega}{\sqrt{\delta_{1}}}\sqrt{\delta_{2}+\log(\delta\omega^{2})}.
\end{equation}
Expanding for small $\omega$,
\begin{equation}
k(\omega)=\pm\frac{\omega}{\sqrt{\delta_{1}}}\sqrt{\delta_{3}\log(\delta\omega^{2})}
-\frac{\omega\delta_{2}}{2\sqrt{\delta_{1}}}(\delta_{3}\log(\delta\omega^{2}))^{-\frac{1}{2}}
+\mathcal{O}(\omega\log^{-3/2}(\delta\omega^{2})).
\end{equation}
It can be seen that the dispersion relation differs from the holographic zero sound mode by logarithmic terms.
\section{A numerical survey of Fermi surface}
In previous section we observed a sound-like excitation in the regime of $n<0$.
It is known that such zero sound mode is associated with the deformation of the Fermi
surface away from the spherical shape. The theory of normal Fermi liquids
tells us that the jump in the distribution function can be observed
as a singularity in the retarded current-current Green's function in the
$\omega=0$ limit. To see if we can observe such Fermi surface we
need the complete solution to the equation for the gauge fluctuations~(\ref{4eq6}) with $\omega=0$,
at least numerically.

We investigated thermodynamics of probe D-branes in section 3, where we found that when
$\beta\neq2$, the specific heat was proportional to the temperature $T$ under certain conditions,
which is just the behavior of Fermi liquids. As we have many groups of parameters $(\alpha,\eta)$ which
lead to specific heat linear in $T$, we choose the following parameters $\alpha=3/2$,$\eta=1/2$ in $D=4$
dimensional spacetime as an example. For simplicity we also fix $q=1$ and $V_0=1$. Then
the equation for the gauge fluctuations~(\ref{4eq6}) in the $\omega=0$ limit is reduced to
\begin{equation}    \label{eq:red}
E^{\prime\prime} + \frac{3(r^{3} - 2d^{2})}{2r(r^{3} + d^{2})}E^{\prime}
-\frac{15k^{2} \sqrt{r}}{8(r^{3} + d^{2})} E=0.
\end{equation}
In the near horizon region $r \to 0$, the perturbative solution of the above equation is given by
\begin{equation}
E = c_{1} + c_{2} r^{4} .
\end{equation}
Here, we choose the boundary condition $E = 0$ at the horizon, which is compatible to
impose the incoming boundary condition for $\omega \ne 0$. So we set $c_1 = 0$.
At the boundary $r=\infty$, the asymptotic solution of (\ref{eq:red}) becomes
\begin{equation}
E = A + B r^{-1/2} .
\end{equation}
Notice that since the first term is finite at the boundary, it corresponds to the source term
and the coefficient of the second term implies the vev of the dual gauge operator. Then, the
Green function is proportional to $- B/A$. In Figure 1, we plot the dependence of the Green's function
on the momentum $k$ for $d=1$.

\begin{figure}
\begin{center}
\vspace{1cm}
\includegraphics[angle=0,width=0.5\textwidth]{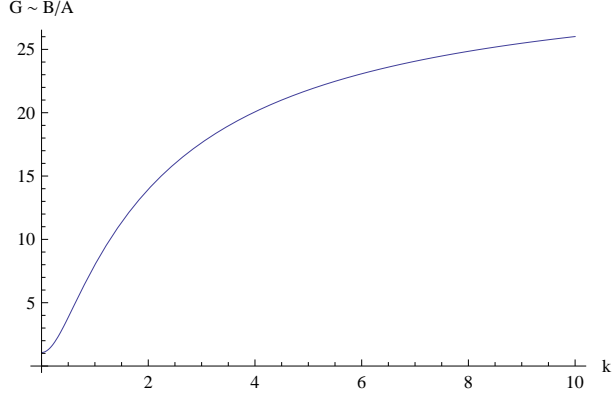}
\vspace{-0cm}
\caption{\small Ratio B/A as a function of $k$.}
\label{number}
\end{center}
\end{figure}
From the figure it can be easily seen that the characteristic structure does not appear
in a wide range of $k$, while the specific heat and zero sound exhibit typical features of
Fermi liquids. Such phenomenon was also observed in~\cite{Kulaxizi:2008jx}. It was pointed
out in~\cite{Kulaxizi:2008jx} that one difficulty in applying Landau's theory of Fermi liquids was
the assumption that the particle number should be conserved as the strength of the interaction was
varied, which was not obvious for the case they discussed. Their conclusion seems to be still applicable to
the present case.
\section{AC conductivity}
In this section we calculate the AC conductivity by making use of the correlation functions, which
can be seen as a by-product of section 4. Furthermore, we will take the limit of zero density
and compare the results with those appeared in previous literatures.

It can be easily seen that the current-current correlation function is given by
\begin{equation}
G^{R}_{xx}(\omega,k=0)=\omega^{2}\Pi(\omega,k=0)
\propto\frac{-\tau_{\rm eff}C_{1}\omega^{2}}{\delta_{2}\omega^{2}+\delta_{3}G_{0}(\omega)}.
\end{equation}
Therefore we can obtain the following expressions in the small frequency limit,
\begin{equation}
G^{R}_{xx}(\omega\rightarrow0,k=0)\propto-\tau_{\rm eff}C_{1}\delta_{3}^{-1}\omega^{\frac{n}{\beta-1}}
~~~n>0,
\end{equation}
\begin{equation}
G^{R}_{xx}(\omega\rightarrow0,k=0)\propto-\frac{\tau_{\rm eff}C_{1}\omega^{2}}{\delta_{3}
\log(\delta\omega^{2})}
~~~n=0,
\end{equation}
\begin{equation}
G^{R}_{xx}(\omega\rightarrow0,k=0)\propto-\tau_{\rm eff}C_{1}\delta_{2}^{-1},
~~~n<0
\end{equation}
Recalling that the definition of AC conductivity is given by
\begin{equation}
\sigma(\omega)=-\frac{i}{\omega}G^{R}_{xx}(\omega\rightarrow0,k=0),
\end{equation}
we can arrive at the following results
\begin{equation}
\sigma(\omega)\propto i\tau_{\rm eff}C_{1}\delta_{3}^{-1}\omega^{\frac{n}{\beta-1}-1},~~~n>0,
\end{equation}
\begin{equation}
\sigma(\omega)\propto i\tau_{\rm eff}C_{1}\delta_{3}^{-1}\omega^{-1}(\log(\delta\omega^{2}))^{-1},~~~n=0,
\end{equation}
\begin{equation}
\sigma(\omega)\propto i\tau_{\rm eff}C_{1}\delta_{2}^{-1}\omega^{-1},~~~n<0.
\end{equation}
As a check of consistency, we consider the specific limit, $\beta=2,\gamma=2/\textsf{z}, n=1-2/\textsf{z}$.
When $n>0,\textsf{z}>2$, $\sigma(\omega)\propto i\tau_{\rm eff}C_{1}\delta_{3}^{-1}\omega^{-2/\textsf{z}}$,
which agrees with the result obtained in~\cite{HoyosBadajoz:2010kd}. The behavior of the AC conductivity
is qualitatively similar to that investigated in~\cite{HoyosBadajoz:2010kd}. To be concrete, when $n\neq0$
$\delta_{2}$ is real while $\delta_{3}$ is complex. The conductivity is purely imaginary when $n<0$ and has
a simple pole at zero frequency. Then according to Kramers-Kronig relation, one can conclude that the real
part of the conductivity, and hence the spectral function, consists of a delta function at zero frequency.
When $n>0$ the conductivity, and hence the spectral function has a power-law dependence, and no explicit
quasi-particle excitation exists.

On the other hand, it was observed in~\cite{Horowitz:2009ij} that after transforming the equation
of $a_{x}$ into a Schr\"{o}dinger-like form, the AC conductivity was
directly related to the reflection amplitude for scattering off the potential.
For our system, it can be observed that the
equation for $a_{x}$ at $k=0$ becomes
\begin{equation}
\partial_{r}[\frac{e^{-\phi}g_{xx}^{q/2-1}|g_{tt}|a_{x}^{\prime}}
{(|g_{tt}|g_{rr}-A^{\prime2}_{t})^{1/2}}]+\frac{e^{-\phi}g^{q/2-1}_{xx}g_{rr}}
{\sqrt{|g_{tt}|g_{rr}-A^{\prime2}_{t}}}\omega^{2}a_{x}=0
\end{equation}
in the original coordinate,
$$ds^{2}=-C_{1}r^{\beta}dt^{2}+\frac{dr^{2}}{C_{1}r^{\beta}}+C_{3}r^{\gamma}
\sum\limits^{D-2}_{i=1}dx^{2}_{i}.$$
The asymptotic behavior of $a_{x}$ is given by
\begin{equation}
a_{x}(\omega)=\frac{E_{x}(\omega)}{i\omega}+\frac{J_{x}(\omega)}{\tau_{\rm eff}(\gamma-\beta-m+1)}r^{\gamma
-\beta-m+1},
\end{equation}
where we have introduced a background electric field $E_{x}(t)\equiv {\rm Re}E_{x}(\omega)e^{-i\omega t}$,
The above equation can be put in a Schr\"{o}dinger-like form,
\begin{equation}
-\frac{d^{2}\Psi}{ds^{2}}+U(s)\Psi=\Omega^{2}\Psi,
\end{equation}
where
\begin{equation}
a_{x}=\frac{r^{\gamma/2}}{\lambda^{1/2}}\Psi,~~~\lambda=\sqrt{r^{2m}+d^{2}},~~~
\frac{d}{dr}=r^{-\beta}\frac{d}{ds},~~~\Omega=\frac{\omega}{C_{1}},
\end{equation}
and
\begin{equation}
U(s)=r^{\gamma/2}\lambda^{-1/2}[\frac{1}{2}\frac{d}{ds}(r^{-\gamma/2}\lambda^{-1/2}\frac{d\lambda}{ds})
-\frac{d}{ds}(\frac{\gamma}{2}r^{-\gamma/2-1}\lambda^{1/2}\frac{dr}{ds})].
\end{equation}

Generally speaking, the AC conductivity can be obtained by numerical methods. However, in the
zero density limit
\begin{equation}
d=0,~~~\lambda=r^{m},
\end{equation}
one can find
\begin{equation}
U(s)=\frac{U_{0}}{s^{2}},~~~U_{0}=\frac{(m-\gamma)(m-\gamma+2\beta-2)}{4(\beta-1)^{2}},
\end{equation}
so the potential possesses the same form as those investigated in~\cite{Goldstein:2009cv}
and~\cite{Chen:2010kn}. Assuming that the solution asymptotes to $AdS_{D}$, we can obtain
\begin{equation}
{\rm Re}(\sigma)\sim\omega^{\kappa},~~~\kappa=2\nu_{0}-1,~~~\nu_{0}^{2}=U_{0}+\frac{1}{4}
\end{equation}
following their approach. In particular, when $q=2,C_{2}=0$, the potential vanishes
and the conductivity is constant at all temperatures
\begin{equation}
\sigma(\omega)=\tau_{\rm eff}\equiv\sigma_{0},
\end{equation}
which agrees with the analysis performed in~\cite{Herzog:2007ij}.
\section{Summary and discussion}
In this paper we explore the zero sound in $D$-dimensional effective holographic theories, whose
bulk fields include the graviton $g_{\mu\nu}$, the $U(1)$ gauge field $\mathcal{A}_{\mu}$
and the scalar field $\phi$. The solutions possess anisotropic scaling symmetry and
they reduce to previously known examples, such as $AdS_{D}$, $AdS_{2}\times\textbf{R}^{D-2}$
and ``modified'' Lifshitz solutions under certain conditions. We consider thermodynamics of massless
probe D-branes in the near-extremal background and clarify the conditions under which the specific
heat is linear in the temperature, which is a characteristic feature of Fermi liquids. Subsequently
we study the zero sound mode by considering the fluctuations of the worldvolume gauge fields on the
probe D-branes. Rather than analytically solving the equations of motion, we obtain the low-frequency
behavior by solving the equation in two different limits and then matching the two solutions in
a regime where the limits overlap, following~\cite{Karch:2008fa} and~\cite{HoyosBadajoz:2010kd}.
The resulting behavior of the zero sound looks similar to that investigated in~\cite{HoyosBadajoz:2010kd},
that is, when the parameter $n\equiv\beta-\gamma-1<0$, the dispersion relation reveals a quasi-particle
excitation; while the zero sound is not a well-defined quasi-particle when $n\geq0$. Furthermore, we
plot the correlation function in $D=4$ at $\omega=0$ with fixed parameters which lead to linear specific
heat. The result is that we cannot observe any characteristic structure of Fermi liquids in a wide range of
$k$, which is similar to what was found in~\cite{Kulaxizi:2008jx}. As a by-product, we also evaluate the
AC conductivity via the current-current correlation function, which reduces to previously known results
at specific limits.

By now there are mainly two approaches for studying condensed matter physics in the context of holography.
One approach can be thought of as ``top-down'', that is, we consider certain exact solutions or brane
configurations in string/M theory which possess the desired properties of condensed matter systems. The main
advantage is that we have clear understanding about the dual field theories, while it is difficult to
find such solutions in string/M theory. A complementary approach can be seen as ``bottom-up'', which means that
we consider certain toy models of gravity which possess solutions with the desired properties. It allows a
parametrization of large classes of IR dynamics and provides useful information in the dual field theory
side. However, the main disadvantage is that the embeddings of such toy models into string/M theory are not
obvious, thus many things in the field theory side remain unknown. However, the ``bottom-up'' approach
is an efficient tool for investigating the AdS/CMT correspondence.

The background solutions we studied in this paper are exact solutions of theories with domain wall vacua.
In~\cite{Perlmutter:2010qu} the author also constructed interpolates between two exact solutions of the single-
exponential, domain wall gravity theory, which lead to the argument that the domain wall/QFT correspondence
~\cite{Boonstra:1998mp} can be considered as an effective holographic tool which is applicable in settings
beyond the regime of domain wall supergravities. Therefore the domain wall/QFT correspondence can also be
taken as one specific class of effective holographic theories. Moreover, the author of~\cite{Perlmutter:2010qu}
argued that even when the UV completion of some bulk theory was unknown, if the theory admitted an approximate
domain wall solution at some intermediate value of $r$ then one could use domain wall/QFT correspondence to
develop a holographic map. Thus it is interesting to develop the AdS/CMT holography by making use of this
domain wall/QFT correspondence. In particular, we can establish precision holography in this
anisotropic background along the line of~\cite{Kanitscheider:2008kd} and study the fermionic correlation
functions following~\cite{Gubser:2009dt, Faulkner:2009am, Faulkner:2010da}. We leave such fascinating
projects in the future.

\bigskip \goodbreak \centerline{\bf Acknowledgments}
\noindent This work was supported by the National Research
Foundation of Korea(NRF) grant funded by the Korea government(MEST)
through the Center for Quantum Spacetime(CQUeST) of Sogang
University with grant number 2005-0049409.

\appendix
\section{Asymptotic expansions of $E(z)$ at specific values}
In this appendix we show that when the parameter $m$ takes some specific
values, the large $z$ expansions of the solutions $E(z)$ to
equation~(\ref{small}) cannot match that of~(\ref{daz}).
Firstly, when $m=1$ but $m\neq-n$, the solution to~(\ref{small})
is given by
\begin{eqnarray}
E(z)&=&D_{1}+D_{2}[C_{1}k^{2}\big(\frac{1}{\sqrt{1+d^{2}z^{2}}}+\log z-\log2(1+\sqrt
{1+d^{2}z^{2}})\big)\nonumber\\& &+\omega^{2}\big(\frac{z^{\beta-\gamma}}{\beta-\gamma}{}_{2}F_{1}(-\frac{1}{2},\frac{\beta-\gamma}{2};
1+\frac{\beta-\gamma}{2},-d^{2}z^{2})\nonumber\\
& &-\frac{d^{2}z^{2+\beta-\gamma}}{2+\beta-\gamma}{}_{2}F_{1}(\frac{1}{2},1+\frac{\beta-\gamma}{2};
2+\frac{\beta-\gamma}{2},-d^{2}z^{2})\big)].
\end{eqnarray}
When performing the expansion, we just focus on the powers of $z$. The Hypergeometric function gives
\begin{equation}
\sim {\rm const}+z^{\beta-\gamma+1}+z^{-2\gamma},
\end{equation}
which does not match that of~(\ref{daz}).

Secondly, when $m\neq1$ but $m=-n\neq0$, the solution is
\begin{eqnarray}
E(z)&=&D_{1}+D_{2}[C_{1}k^{2}\frac{z^{m-1}}{m}\big(\frac{1}{1+d^{2}z^{2m}}+\frac{1}{m-1}{}_{2}F_{1}
(\frac{1}{2},\frac{1}{2}-\frac{1}{2m};\frac{3}{2}-\frac{1}{2m},-d^{2}z^{2m})\big)\nonumber\\
& &+\omega^{2}\big(\frac{1}{1-\beta+\gamma}+(1-\beta+\gamma)\log z-\log(2+2\sqrt{1+d^{2}z^{2+2\gamma-2\beta}})\big)].
\end{eqnarray}
The $\omega^{2}$ term gives
\begin{equation}
\sim\frac{1}{1-\beta+\gamma}-\log2d,
\end{equation}
which does not match that of~(\ref{small}) either. Therefore we just considered the cases
with $m\neq1$ and $m\neq\gamma-\beta+1$ in the main text.


\end{document}